\documentclass[runningheads,fleqn]{cl2emult}

\usepackage{makeidx}  
\usepackage{graphicx} 
\usepackage{subeqnar} 
\usepackage{multicol} 
\usepackage{cropmark} 
\usepackage{phys}     
\makeindex            

\newcommand{\E}{{\mathrm e}}
\newcommand{\I}{{\mathrm i}}
\newcommand{\D}{{\mathrm d}}

\begin{document}
\title*{Mirrorless oscillation based on resonantly enhanced 4-wave mixing:
All-order analytic solutions}
%
%
%
%
\titlerunning{Resonant 4WM: analytic solutions}
%
\author{M.~Fleischhauer\qquad
%
%
%
\institute{Sektion Physik,  
Universit\"at M\"unchen, 
D-80333 M\"unchen, Germany}
e-mail: mfleisch@theorie.physik.uni-muenchen.de}

\maketitle              

\begin{abstract}
The phase transition to mirrorless oscillation in resonantly
enhanced four-wave mixing in double-$\Lambda$ systems
are studied analytically for the ideal case of infinite
lifetimes of ground-state coherences. The stationary susceptibilities 
are obtained in all orders of the 
generated fields and  analytic solutions of the coupled nonlinear
differential equations for the field amplitudes are derived
and discussed.
\end{abstract}


\section{Introduction}


The possibility to cancel the linear absorption in resonant 
atomic systems by means of electromagnetically induced transparency
(EIT) \cite{EIT} lead in recent years to fascinating new developments 
in nonlinear optics \cite{Harris90,Stoicheff90}. 
For example coherently driven, resonant atomic vapors
under conditions of EIT 
allow for complete frequency conversion in distances short 
enough, such that phase matching requirements become irrelevant
\cite{Jain96}. Furthermore the large nonlinearities of these systems 
may lead to a new regime of nonlinear quantum optics on the
few-photon level \cite{Iamamoglu,Hau99b} with potential applications 
to single-photon quantum control \cite{Yamamoto98,Werner_preprint}
and quantum information processing.

One particularly interesting nonlinear process 
based on EIT is the
resonantly enhanced 4-wave mixing in a double-$\Lambda$ system
with counter-propagating pump modes
\cite{Hemmer95}. It has been shown experimentally \cite{zibrov98}
and theoretically \cite{Loeffler98,review} that this system can show a
phase transition to mirrorless oscillations for rather low pump powers.
Close to the threshold of oscillation an almost
perfect suppression of quantum fluctuations
of one quadrature amplitude of a combination mode of the generated 
fields occurs \cite{Yuen79,Lukin99}. Also sufficiently above threshold
light fields with beat-frequencies tightly locked to the
atomic Raman-transition 
and extremely low relative  bandwidth
are generated \cite{Fleischhauer_preprint}. 

All previous studies 
of resonantly enhanced 4-wave mixing
were done in the perturbative regime of small
amplitudes of the generated fields. 
In the present paper I want to discuss
the case of arbitrary amplitudes. 
Using a  simplified open-system model I will derive stationary 
propagation equations for the field amplitudes and present analytic
solutions of these equations. It will be shown that in an ideal case
complete conversion can be achieved within a relatively small
interaction length.


\section{Model and Atomic Polarizations}


I here consider the 
propagation of four electromagnetic waves  in a 
medium consisting of double-$\Lambda$ atoms (see Fig.1).
These waves include two counter-propagating 
driving fields with equal frequencies 
$\nu_{\rm d}$ and Rabi-frequencies $\Omega_{1}$ and $\Omega_{2}$, 
and two probe fields (anti-Stokes and Stokes)  
 described by the
complex Rabi-frequencies $E_1$ and $E_2$,
with carrier frequencies 
$\nu_1 = \nu_{\rm d}+
\omega_0$ and $\nu_2=\nu_{\rm d}-\omega_0$, 
where $\omega_0=\omega_{\rm b1}-
\omega_{\rm b2}$ is the ground-state frequency splitting. The
fields interact via the long-lived coherence on the
dipole-forbidden transition between the metastable ground states 
$b_1$ and $b_2$. 
We assume
that the driving field $\Omega_1$ is in resonance with the 
$b_2\to a_1$ transition, whereas the second driving field
$\Omega_2$ has a detuning $\Delta\gg |\Omega_2|$ from the 
$b_1\to a_2$ transition. In this case linear losses of the fields 
due to single-photon absorption processes are minimized.

\begin{figure}
\sidecaption
\includegraphics[width=.4\textwidth]{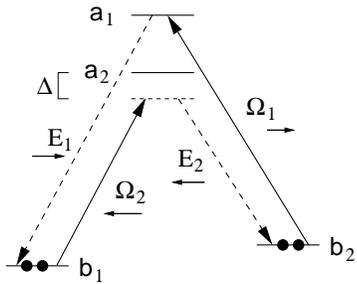}
\caption[]{Atoms in double $\Lambda$ configuration interacting 
with two driving fields 
($\Omega_{1,2}$) and two generated fields ($E_{1,2}$)}
\label{eps1}
\end{figure}

Due to coherent Raman-scattering
the pump fields generate counter-propagating anti-Stokes and Stokes fields. 
For a sufficiently large
density-length product of the medium and for a certain 
pump field intensity, the system shows a phase-transition
to self-oscillations \cite{zibrov98}. 
The feedback mechanism required for an oscillation
is provided here by the 
gain medium: A Stokes photon spontaneously generated
on the $a_2\to b_2$ transition
propagates in the $-z$ direction and
stimulates the generation of an anti-Stokes
photon. This anti-Stokes photon has a different frequency 
but a fixed relative phase and propagates in the $+z$ direction.
It stimulates the generation of another Stokes photon upstream.
The second Stokes photon will be in phase with the first one,
provided that the system is approximatly
phase matched and that there has been no decay of the 
Raman coherence. The phase-locked emission of the second Stokes
photon then closes the feedback loop.
We have shown in \cite{Fleischhauer_preprint} that
phase-matching enforces a strong pulling of the 
beat-note of generated and pump fields to
the atomic Raman transition. I will therefore assume here that
both $\Lambda$ systems are in perfect two-photon resonance.

In order to calculate the medium response to the fields,
one would have to solve the atomic density matrix equations
to all orders in all fields taking into
account all relaxation rates. Although this is in principle possible
it leads to extremely involved expressions. Instead I here use
a simplified open-system model which allows to derive rather
compact expressions for the atomic susceptibilities. 

Since the effects of spontaneous
emission are negligible in the present system, we may model
all relaxations out of the excited states $a_1$
and $a_2$ by rates $\gamma$ out of the system. In thermal equilibrium,
i.e. in the absence of all fields, both lower states $b_1$ and $b_2$
are equally populated. I therefore assume --  within the
open-system approach -- that the atoms are pumped 
 into states $b_1$ or $b_2$ with
 50\% probability respectively.
The corresponding rate is denoted as $r$ and will later be determined
by the requirement that the total probability to find an atom in any of the
states is unity. The finite lifetime of the lower-level coherence
will here be described by a decay out of all states with rate
$\gamma_0$. Thus the open-system model corresponds to the 
experimentally relevant situation of an atomic beam or 
a finite-temperature vapor with  time-of-flight broadening. In this case 
the system can be described by  generalized Schr\"odinger-equations for
field amplitudes instead of density-matrix equations. 

The interaction Hamiltonian of an atom at position $z$ with
the fields can be written in the form
\begin{eqnarray}
H_{\rm int}&=&-\hbar\Bigl[\Omega_1(z)
\,\E^{-\I\nu_\D t}
 |a_1\rangle\langle b_2| +\Omega_2(z) \,
 \E^{-\I\nu_\D t}|a_2\rangle\langle b_1| +\nonumber\\
&& \quad + E_1(z)\, \E^{-\I\nu_1 t} |a_1\rangle\langle b_1| 
+ E_2(z)\, \E^{-\I\nu_2 t} |a_2\rangle\langle b_2| +{\rm adj.}
\Bigr].
\end{eqnarray}
If we denote the state vector of the atom as
\begin{equation}
|\Psi\rangle = a_1 \E^{-\I\nu_{a1} t} |a_1\rangle +
a_2 \E^{-\I(\nu_{a2}-\Delta) t} |a_2\rangle + b_1 
\E^{-\I\nu_{b1} t} |b_1\rangle
+b_2 \E^{-\I\nu_{b2} t} |b_2\rangle,
\end{equation}
where $\hbar\nu_\mu$ are the energies of the corresponding states,
we find the following equations of motion 
of the slowly-varying state amplitudes
for an atom at position $z$
\begin{eqnarray}
\dot a_1 &=& -\left(\gamma_0+\gamma\right) a_1 +\I\Omega_1
 b_2 + \I E_1 b_1,\label{a1}\\
\dot a_2 &=& -\left(\gamma_0+\gamma+i\Delta\right)a_2 + 
\I\Omega_2 b_1 +\I E_2 b_2,\\
\dot b_1 &=& r_1-\gamma_0 b_1+\I\Omega_2^* a_2 +\I E_1^* a_1,\\
\dot b_2 &=& r_2-\gamma_0 b_2 +\I\Omega_1^* a_1 +\I E_2^*
 a_2.\label{b2}
\end{eqnarray}
Here I have introduced 
the rates $r_1$ and $r_2$ to distinguish the cases of pumping into $b_1$
($r_1=r, r_2=0$) and into $b_2$ ($r_1=0, r_2=r$). Note that 
simultaneously setting
$r_1=r_2=r$ corresponds to a {\it coherent} preparation of
the atoms in a 50--50 superposition of $b_1$ and $b_2$. In order
to describe an {\it incoherent} preparation in these
levels one has to consider the two cases separately and add the
density matrix elements following from both cases.

Solving (\ref{a1}--\ref{b2}) in steady state for the case of
injection into $b_1$, i.e.
for $r_1=r$ and $r_2=0$ one finds
\begin{eqnarray}
a_1^{(1)} &=&-\I r\frac{\Omega_1\Omega_2 E_2^*
-E_1|E_2|^2}
{|\Omega_1\Omega_2-E_1 E_2|^2},\\
a_2^{(1)} &=&\I r \frac{|\Omega_1|^2 
\Omega_2
-\Omega_1^* E_1 E_2 
}
{|\Omega_1\Omega_2-E_1 E_2|^2},\\
b_1^{(1)} &=& \I r \frac{\Delta |\Omega_1|^2}
{|\Omega_1\Omega_2-E_1 E_2|^2},\\
b_2^{(1)} &=& -\I r \frac{\Delta \Omega_1^*E_1 }
{|\Omega_1\Omega_2-E_1 E_2|^2},
\end{eqnarray}
where I have used that $\Delta\gg \gamma\gg\gamma_0$ and have kept only
the leading terms.
Similarly one finds for injection into $b_2$, i.e.
for $r_1=0$ and $r_2=r$:
\begin{eqnarray}
a_1^{(2)} &=& \I r\frac{\Omega_1|\Omega_2|^2 
-E_1E_2 \Omega_2^*}                                  
{|\Omega_1\Omega_2-E_1 E_2|^2},\\
a_2^{(2)} &=&-\I r \frac{\Omega_1 \Omega_2 E_1^*
- |E_1|^2 E_2}
{|\Omega_1\Omega_2-E_1 E_2|^2},\\
b_1^{(2)} &=& -\I r \frac{\Delta \Omega_1 E_1^*
}
{|\Omega_1\Omega_2-E_1 E_2|^2},\\
b_2^{(2)} &=& \I r \frac{\Delta |E_1|^2}
{|\Omega_1\Omega_2-E_1 E_2|^2}.
\end{eqnarray}
Taking into account only the leading order contribution in the
above expressions is essentially equivalent to assuming an infinitely
long lived ground-state coherence between $b_1$ and $b_2$.
In vapor cells with coated walls or by using buffer gases,
lifetimes of Hyperfine coherences in alkali vapors in the millisecond
regime are possible. Hence neglecting contributions from finite values
of $\gamma_0$ seems justified. However, in this case also linear
absorption losses are neglected. As a consequence the threshold condition
becomes independent on the pump intensity and
an arbitrarily small flux of pump photons is sufficient to maintain 
oscillations \cite{Fleischhauer_preprint}. 
If on the other hand a small but finite
ground-state dephasing rate is taken into account, the threshold
condition does depend on the pump intensity leading to a lower limit
of the pump-photon flux. In the present paper I am interested 
only in the analytic behavior of the fields in the ideal limit and therefore
the small but finite linear losses associated with the ground-state
dephasing will be ignored.

The pump rate $r$ can be  determined from the
 normalization condition 
$ \sum_\mu  \varrho_{\mu\mu}^{(1)} + \varrho_{\mu\mu}^{(2)} = 1$.
%
One finds $ r=\bigl(|\Omega_1\Omega_2-E_1 E_2|^2\bigr)
/\bigl[\Delta \bigl(|\Omega_1|^2 +|E_1|^2\bigr)\bigr].$
%
With this one obtains for the non-diagonal density matrix elements
$\varrho_{a_\mu b_\nu}= a_\mu^{(1)}b_\nu^{(1)* }+ a_\mu^{(2)}b_\nu^{(2)*}$:
\begin{eqnarray}
\varrho_{a_1b_1} &=&
-\frac{|\Omega_1|^2 \Omega_1\Omega_2 E_2^*
-E_1^2 E_2 \Omega_1^*\Omega_2^*}
{\Delta \left(|\Omega_1|^2+|E_1|^2\right)^2}
-\frac{|\Omega_1|^2\left(|\Omega_2|^2-|E_2|^2\right)}
{\Delta \left(|\Omega_1|^2+|E_1|^2\right)^2}\, E_1,\\
\varrho_{a_1b_2} &=&
\frac{\Omega_1^2\Omega_2 E_1^*E_2^*-|E_1|^2E_1 E_2\Omega_2^*}
{\Delta \left(|\Omega_1|^2+|E_1|^2\right)^2}
+\frac{|E_1|^2(|\Omega_2|^2+|E_2|^2)}
{\Delta \left(|\Omega_1|^2+|E_1|^2\right)^2}\,\Omega_1,\\
\varrho_{a_2b_1} &=&
-\frac{\left(|\Omega_1|^2+|E_1|^2\right)E_1 E_2\Omega_1^*}
{\Delta \left(|\Omega_1|^2+|E_1|^2\right)^2}
+\frac{|\Omega_1|^2\left(|\Omega_1|^2+|E_1|^2\right)}
{\Delta \left(|\Omega_1|^2+|E_1|^2\right)^2}\, \Omega_2,\\
\varrho_{a_2 b_2} &=&-\frac{\left(|\Omega_1|^2+|E_1|^2\right)\Omega_1\Omega_2
E_1^*}
{\Delta \left(|\Omega_1|^2+|E_1|^2\right)^2}
+\frac{
|E_1|^2\left(|\Omega_1|^2+|E_1|^2\right)}
{\Delta \left(|\Omega_1|^2+|E_1|^2\right)^2}\, E_2.
\end{eqnarray}
The first terms in these expressions describe the
nonlinear coupling between the modes and the second ones
ac-Stark shift induced changes in the refractive indices.
It should be noted that there are no imaginary linear
susceptibilities, i.e. there is no linear dissipation despite
the fact, that $\Omega_1$ and $E_1$ are in
single-photon resonance.


\section{Stationary field equations and analytic solutions}


In slowly-varying amplitude and phase approximation, the field amplitudes
obey the following equation of motion
\begin{eqnarray}
\frac\D{\D z} E_1 &=& \I k_1 E_1 + \I\frac{\wp^2 k_1}
{2\hbar\varepsilon_0} N\, \varrho_{a_1b_1},\\
\frac\D {\D z} E_2^* &=& \I k_2 E_2^* + \I\frac{\wp^2 k_2}
{2\hbar\varepsilon_0} N\, \varrho_{a_2b_2}^*,\\
\frac\D {\D z} \Omega_1 &=& \I k_\D  \Omega_1 + \I
\frac{\wp^2 k_\D }
{2\hbar\varepsilon_0} N\, \varrho_{a_1b_2},\\
\frac\D {\D z} \Omega_2^* &=& \I k_\D  \Omega_2^* 
+ \I\frac{\wp^2 k_\D }
{2\hbar\varepsilon_0} N\, \varrho_{a_2b_1}^*,
\end{eqnarray}
where $k_1$, $k_2$ and $k_\D $ are the free-space wavenumbers of
the generated and pump fields, $N$ is the atomic number density
and $\wp$ are the dipole moments of the
corresponding transitions, which have been assumed to be equal for simplicity.
Since the wavenumbers of the fields differ only slightly, one may
approximate the coupling parameter in all equations by $\kappa\equiv
\wp^2k_\D  N/2\hbar\varepsilon_0$.
Introducing field amplitudes which are slowly varying in space,
$E_1=\widetilde E_1 \, \E^{\I k_1 z}, 
E_2=\widetilde E_2 \, \E^{-\I k_2 z}, \Omega_1=\widetilde \Omega_1
\, \E^{\I k_\D  z}$ and $\Omega_2=\widetilde \Omega_2\, 
\E^{-\I k_\D  z}$ one eventually arrives at
\begin{eqnarray}
\frac\D {\D z} E_1 &=& -\I\kappa
\frac{|\Omega_1|^2 \Omega_1\Omega_2 E_2^*
-E_1^2 E_2 \Omega_1^*\Omega_2^*}{\Delta \left(|\Omega_1|^2+|E_1|^2\right)^2}
\nonumber\\
&&-\I\left[\Delta k+\kappa
\frac{|\Omega_1|^2\left(|\Omega_2|^2-|E_2|^2\right)}
{\Delta \left(|\Omega_1|^2+|E_1|^2\right)^2}\right]\, E_1,\label{E1}\\
\frac\D {\D z} E_2^* &=& -\I\kappa 
\frac{\left(|\Omega_1|^2+|E_1|^2\right)\Omega_1^*\Omega_2^*
E_1}
{\Delta \left(|\Omega_1|^2+|E_1|^2\right)^2}
\nonumber\\
&&+\I\kappa\frac{
|E_1|^2\left(|\Omega_1|^2+|E_1|^2\right)}
{\Delta \left(|\Omega_1|^2+|E_1|^2\right)^2}\, E_2^*,\\
\frac\D {\D z} \Omega_1 &=&\I\kappa 
\frac{\Omega_1^2\Omega_2 E_1^*E_2^*-|E_1|^2E_1 E_2\Omega_2^*}
{\Delta \left(|\Omega_1|^2+|E_1|^2\right)^2}
\nonumber\\
&&+\I\kappa\frac{|E_1|^2(|\Omega_2|^2+|E_2|^2)}
{\Delta \left(|\Omega_1|^2+|E_1|^2\right)^2}\,\Omega_1\\
\frac\D {\D z} \Omega_2^* &=& -\I\kappa
\frac{\left(|\Omega_1|^2+|E_1|^2\right)E_1^* E_2^*\Omega_1}
{\Delta \left(|\Omega_1|^2+|E_1|^2\right)^2}
\nonumber\\
&&+\I\kappa\frac{|\Omega_1|^2\left(|\Omega_1|^2+|E_1|^2\right)}
{\Delta \left(|\Omega_1|^2+|E_1|^2\right)^2}\, \Omega_2^*,\label{O2}
\end{eqnarray}
where I have dropped the tildes again for notational simplicity, and
$\Delta k=k_2-k_1$ is the free-space phase mismatch. 
Expanding these expressions into third order of the generated
fields $E_1$ and $E_2$ reproduces the equations of 
\cite{Fleischhauer_preprint}.
Equations (\ref{E1}--\ref{O2}) together with the boundary-conditions
\begin{eqnarray}
E_1(0)=0,\quad E_2(L)=0,\quad \Omega_1(0)=\Omega_{10}, \quad
{\rm and} \quad\Omega_2(L)=\Omega_{20},
\end{eqnarray}
where $L$ is the length of the interaction region and 
$\Omega_{10}$ and $\Omega_{20}$ are the given input amplitudes, form
a nonlinear boundary-value problem.
One easily verifies that the set of differential
equations has always the trivial solution $E_1\equiv E_2\equiv 0$, and
$\Omega_1(z)\equiv \Omega_{10}$ and $\Omega_2(z)\equiv \Omega_{20}$.

As has been
discussed in detail in \cite{Fleischhauer_preprint}, the phase mismatch
is easily compensated in an optically dense vapor by a small
detuning from the two-photon resonance. 
Oscillation occurs at frequencies such that the phase-matching
condition is automatically fulfilled.
I therefore set this term equal to zero
in the following. 


\subsubsection{Constants of Motion:}


The field equations have the following constants of motion.
From the energy-momentum conservation follow the Manley-Rowe
relations
\begin{eqnarray}
\frac\D {\D z} \left(|\Omega_1|^2+|E_1|^2\right) &=& 0,\\
\frac\D {\D z} \left(|\Omega_2|^2+|E_2|^2\right) &=& 0,
\end{eqnarray}
which state that each photon taken out of the pump fields
$\Omega_1$ or $\Omega_2$
is put into the anti-Stokes and Stokes fields $E_1$ and $E_2$
respectively. Furthermore one finds that the total intensity of the
pump field is constant in space
\begin{eqnarray}
\frac\D {\D z} \left(|\Omega_1|^2+|\Omega_2|^2\right) = 0.
\end{eqnarray}
The same is true for the generated fields, which however follow already
from the above constants of motion.
\begin{eqnarray}
\frac\D {\D z} \left(|E_1|^2+|E_2|^2\right) = 0.
\end{eqnarray}
Without the phase terms in (\ref{E1}--\ref{O2}), which 
represent contributions due to ac-Stark 
shifts, also the quartic expression 
${\rm Re}\,[\Omega_1\Omega_2 E_1^* E_2^*]$ would be a constant of motion.
In fact the boundary conditions for the generated fields imply
that ${\rm Re}\,[\Omega_1\Omega_2 E_1^* E_2^*]\equiv 0$.
It will be shown later on that 
${\rm Re}\,[\Omega_1\Omega_2 E_1^* E_2^*]$ is in any case to a very good
approximation a constant of motion.


\subsubsection{Amplitude-Phase Equations:}


It is convenient to rewrite the field equations in terms of 
amplitudes and phases. Introducing $E_n=e_n\, \E^{-\I\phi_n}$ and $\Omega_n=
a_n\, \E^{-\I\psi_n}$ ($n=1,2$) one obtains
\begin{eqnarray}
\frac\D {\D z} e_1 &=& \frac{\kappa}{\Delta} 
\frac{a_1 a_2 e_2}{a_1^2+e_1^2}\, \sin\psi,\label{e1}\\
-\frac\D {\D z} e_2 &=& \frac{\kappa}{\Delta} 
\frac{a_1 a_2 e_1}{a_1^2+e_1^2}\, \sin\psi,\\
-\frac\D {\D z} a_1 &=& \frac{\kappa}{\Delta} 
\frac{a_2 e_1 e_2}{a_1^2+e_1^2}\, \sin\psi,\\
\frac\D {\D z} a_2 &=& \frac{\kappa}{\Delta} 
\frac{a_1 e_1 e_2}{a_1^2+e_1^2}\, \sin\psi,
\end{eqnarray}
where $\psi=\phi_1+\phi_2-\psi_1-\psi_2$ is the relative phase between
the fields. It obeys the equation
\begin{eqnarray}
\frac\D {\D z} \psi &=&
\frac{\kappa}{\Delta}
\biggl[\frac{a_1a_2 e_2(a_1^2-e_1^2)}{e_1(a_1^2+e_1^2)^2}
-\frac{a_1a_2 e_1}{e_2(a_1^2+e_1^2)} +
\frac{a_2 e_1 e_2(a_1^2-e_1^2)}{a_1(a_1^2+e_1^2)^2}
\nonumber\\
&&\qquad -\frac{a_1 e_1 e_2}{a_2(a_1^2+e_1^2)}\biggr]\cos\psi
+\frac{\kappa}{\Delta}\biggl[\frac{e_1^4-a_1^4 + 2 e_1^2 a_2^2}{
(a_1^2+e_1^2)^2}\biggr].\label{psi}
\end{eqnarray}
%


\subsubsection{Solution for Equal Input Intensities:}


Let me now consider the case of equal input intensities of both pump
fields, i.e. $a_1(0)=a_{10}=a_{20}=a_2(L)$. Making use of the constants
of motion one can write
\begin{eqnarray}
e_1(z) &=& e\, \sin\vartheta(z),\qquad
a_1(z)=\sqrt{a_{10}^2-e^2\sin^2\vartheta(z)},\\
e_2(z) &=& e\, \cos\vartheta(z),\qquad 
a_2(z)=\sqrt{a_{10}^2-e^2\cos^2\vartheta(z)},
\end{eqnarray}
with the output amplitude of the generated fields $e$
and the mixing angle $\vartheta(z)$ as the only remaining
variables. The boundary conditions are now $\vartheta(0)=0$
and $\vartheta(L)=\pi/2$, if $e\ne 0$, i.e. for the non-trivial
solutions.

Substituting the above expressions into (\ref{e1})
yields the nonlinear equation 
\begin{eqnarray}
\frac\D {\D z} \vartheta(z) =\frac{\kappa}{\Delta}
\left[1-\varepsilon^2+\frac{\varepsilon^4}{4}\sin^2\bigl[2\vartheta(z)\bigr]
\right]^{1/2}\!\!
\sin\psi(z),\label{theta}
\end{eqnarray}
where $\varepsilon\equiv e/a_{10}$.
In order to solve (\ref{theta}) one can in principle introduce a nonlinear
stretch of the spatial coordinate according to
\begin{equation}
\xi(z)=\int_0^z\!\!\D z'\, \sin\psi(z'),\qquad{\rm and}\quad 
\D \xi=\sin\psi(z)\, \D z
\end{equation}
which removes the term $\sin\psi(z)$ on the r.h.s. of
(\ref{theta}). I will show later on, however,
that to a very good approximation $\sin\psi(z)\equiv 1$.
Thus $\xi=z$ and $\sin\psi(z)=1$ is used in
the following. 

Integrating (\ref{theta}) from $z=0$ to $z=L$ leads to an equation for the
normalized output amplitude $\varepsilon=e/a_{10}$:
\begin{eqnarray}
{K\left[\displaystyle\frac{\varepsilon^4}{4(\varepsilon^2-1)}\right]}
 = \frac{\kappa L}{\Delta}\, {\sqrt{1-\varepsilon^2}},\label{e}
\end{eqnarray}
where  $K$ is the complete elliptic
integral of the first kind \cite{Abramovitz}. One easily verifies that
(\ref{e}) has only a real-valued solution $\varepsilon$, if
$\kappa L/\Delta\ge \pi/2$, which is the threshold condition
for mirrorless oscillations \cite{zibrov98,review}. 
For smaller values of $\kappa L/\Delta$
the equations of motion have only the trivial solution.
\begin{figure}
\includegraphics[width=.6\textwidth]{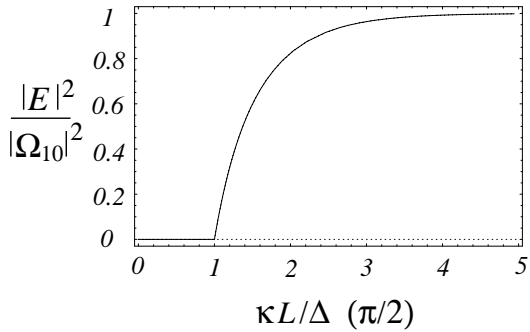}
\caption[]{Output intensity of generated fields 
$E\equiv E_1(L)=E_2(0)$
normalized to input intensity of pump fields
as function
of effective interaction length, $\Omega_1(0)=\Omega_2(L)\equiv\Omega_{10}$}
\label{eps2}
\end{figure}
Figure 2 shows the output intensity of the generated fields
normalized to the input intensity of the pump fields as a function of
the effective density length product $\kappa L$.
One clearly recognizes that for a sufficiently large 
product $\kappa L$ complete conversion can be achieved.

The spatial behavior  of the field strength inside the
vapor cell can be obtained from incomplete elliptical
integrals following from (\ref{theta}). 
Figure 3 shows the field amplitudes inside the medium
for $\varepsilon=0.2$, i.e. just above threshold and
for $\varepsilon=0.98$ i.e. for almost complete conversion.
\begin{figure}
\includegraphics[width=1.0\textwidth]{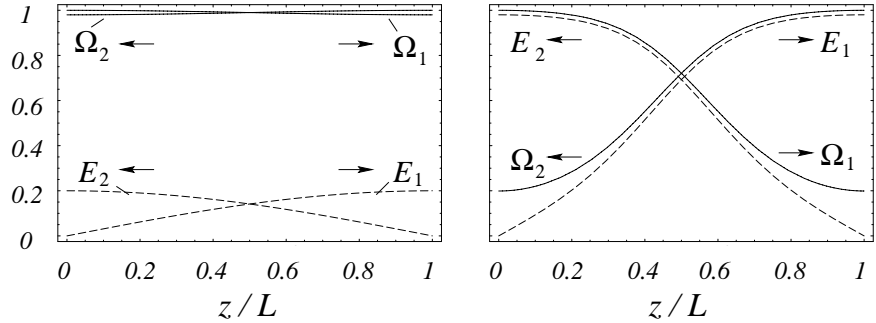}
\caption[]{Field amplitudes inside interaction region
for small conversion  $\varepsilon=E/\Omega_{10}=0.2$ (left) and large
conversion $\varepsilon=E/\Omega_{10}=0.98$ (right)}
\label{eps3}
\end{figure}

Not to far above threshold, the square root in (\ref{theta})
can be expanded and one  recovers the third-order solution obtained in 
\cite{Fleischhauer_preprint}: 
\begin{eqnarray}
\vartheta(z) \approx \frac{\kappa z}{\Delta}\left(1-\frac{1}{2}
\varepsilon^2\right)
\end{eqnarray}
with
\begin{eqnarray}
\varepsilon = \sqrt{2}\left[1-\frac{\pi}{2}\frac{\Delta}{\kappa
L}\right]^{1/2}
\qquad{\rm for}\quad\frac{\kappa L}{\Delta}\ge 
\frac{\pi}{2}.
\end{eqnarray}

In order to verify the approximation $\sin\psi(z)\equiv 1$, I have numerically
integrated the differential equation (\ref{psi}) with the above
solutions. Figure 4 shows the comparison between the nonlinear coordinate
$\xi(z)$ and $z$ for the case $\varepsilon=0.98$. One recognizes that
$\xi$ deviates from $z$ by at most 1\%. For smaller
conversions an even smaller difference shows up. Therefore the
approximation $\sin\psi=1$ is very well justified. This also implies
that ${\rm Re}\, [\Omega_1\Omega_2 E_1^* E_2^*]$ is to a very good 
approximation a constant of motion.
\begin{figure}
\includegraphics[width=.55\textwidth]{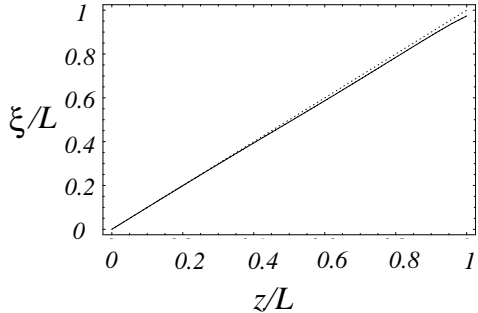}
\caption[]{Effective interaction distance $\xi$
versus physical interaction distance $z$ for
large conversion ($\varepsilon=E/\Omega_{10}=0.98$). Dotted line
corresponds to  $\xi=z$}
\label{eps4}
\end{figure}


\section{Summary}


In the present paper all-order atomic susceptibilities for
resonantly enhanced 4-wave mixing are presented and field equations
derived. The coupled nonlinear differential equations are solved analytically
for the case of infinitely long-lived ground-state coherences
and under the assumption of negligible phase changes due to
ac-Stark shifts. Below a certain critical value of the density-length
product only the trivial solution exists, where the generated Stokes and
anti-Stokes components have vanishing amplitude. Above the threshold
to mirrorless oscillations the photon conversion efficiency
increases very rapidly and at a density-length product of about 3
times the threshold value, 95\% conversion is achieved.


\section*{Acknowledgement}


The author would like to thank the organizers of the International Conference
on Laser Physics and Quantum Optics, ICLPQO'99, in particular
Prof. Shi-Yao Zhu for the invitation and the  hospitality in China. 
The financial support of the German Science Foundation is gratefully
acknowledged.



\clearpage
\addcontentsline{toc}{section}{Index}
\flushbottom
\printindex

\end{document}